\documentclass[reprint,pra]{revtex4-1}
\usepackage{graphicx}
\usepackage{color}
 \usepackage{bm}
\usepackage{amsmath}
\usepackage{amssymb}

\begin{document}

\title{Resonant field enhancement in lossy periodic structures
  supporting \\ complex bound states in the continuum}

\author{Ling Tan}
\author{Lijun Yuan}
\email{ljyuan@ctbu.edu.cn}
\affiliation{College of Mathematics and Statistics, Chongqing Technology and Business University, Chongqing, 
China \\ Chongqing Key Laboratory of Social Economic and Applied Statistics, Chongqing Technology and Business University, Chongqing, China}

\author{Ya Yan Lu}
\affiliation{Department of Mathematics, City University of Hong Kong, 
  Kowloon, Hong Kong, China} 
\date{\today}

\begin{abstract}
Resonant modes in a lossy periodic structure sandwiched between two
lossless homogeneous media form bands that depend on the Bloch
wavevector continuously and have a complex frequency due to radiation and
absorption losses. A complex bound state in the continuum (cBIC)
is a special state with a zero radiation loss in such a band. 
Plane waves incident upon the periodic structure induce local
fields that are resonantly enhanced. 
In this paper, we derive a rigorous formula for field enhancement, and
analyze its dependence on the frequency, wavevector and amplitude of
the incident wave. For resonances with multiple radiation channels, we
determine the incident wave that maximizes the field enhancement,  and
find conditions under which the field enhancement can be related to the
radiation and dissipation quality factors. We also show that with
respect to the Bloch wavevector, the largest field enhancement is
obtained approximately when the radiation and dissipation quality
factors are equal. Our study clarifies the various factors related to field enhancement,
and provides a useful guideline for applications where a strong local 
field is important. 
\end{abstract}
\maketitle 

\section{Introduction}

A strong local field is important to many applications in photonics,
such as sensing~\cite{homola99,fan08,zhang15,romano18}  
and imaging~\cite{yesi19,roma20}, and is essential to the enhnancement
of emissive processes and nonlinear optical effects~\cite{maier,kim08, yuan16,
  yuan17, kosh19,kosh20,yuan20}. For structures supporting a high quality factor ($Q$ factor)
resonance, if the frequency of the incident wave is close to the
resonant frequency, the amplitude of the local field can be much
larger than that of the incident wave~\cite{yoon15,moca15,bulg17,hu20_1,hsu21,bin21}.
The strength of resonant field enhancement, i.e. the ratio of the
amplitudes of the local field and the incident wave, depends on the
resonant mode and its coupling with the incident
wave~\cite{maier,yoon15,hu20_1}. 
For unbounded structures that are periodic or invariant in at least
one spatial direction, resonant modes form bands that vary with the
wavevector continuously. In recent years, dielectric periodic
structures supporting bound states in the continuum (BICs) have been
intensively
studied~\cite{hsu13_2,yang14,bulg14,hu15,yuan17_1,jin19,
  fudan,song20,yoda20}. 
In a lossless periodic structure, a BIC is a special bound state 
in a  band of resonant modes. As the
wavevector of the resonant modes tends to that of the BIC, the $Q$
factor diverges~\cite{hsu13_2}, and in principle, the field enhancement can be
arbitrarily large~\cite{hu20_1}.   

However, electromagnetic waves in practical materials always suffer
from absorption loss (dissipation). For a lossy periodic  structure surrounded by
homogeneous and lossless dielectric media, any eigenmode with a real
wavevector must have a complex frequency.
A bound state is an eigenmode that decays exponentially in the surrounding 
homogeneous media. If the real part of its complex frequency lies in
the radiation continuum, the bound state will be
referred to as a complex BIC (cBIC) in this paper.  
For periodic structures with an in-plane inversion symmetry,
symmetry-protected cBICs are easy to find~\cite{hu20_2}.
A resonant mode is an eigenmode that radiates out power in the
surrounding media.
In a lossy structure, the total $Q$ factor ($Q_{\rm tot}$) of a
resonant mode satisfies $1/Q_{\rm tot} = 1/Q_{\rm rad} + 1/Q_{\rm
  dis}$, where $Q_{\rm rad}$ and $Q_{\rm dis}$ are the $Q$ factors
associated with radiation and absorption losses, 
respectively~\cite{blio08}.
In a lossy periodic structure, if there is a cBIC in a band
of resonant modes, then $Q_{\rm rad} \to \infty$ and $Q_{\rm dis}$
remains bounded, as the wavevector tends to that of the cBIC.
It has been shown that field enhancement is 
related to $Q_{\rm rad}$ and $Q_{\rm dis}$, and the maximum field
enhancement is obtained at the wavevector 
satisfying $Q_{\rm rad} = Q_{\rm dis}$ (the critical coupling
condition)~\cite{yariv02,yoon15}.

The existing results on resonant field enhancement are derived from
the temporal coupled mode theory (CMT)~\cite{maier,yoon15}. The CMT is a powerful
modeling technique that allows one to reveal the most important physical
phenomena. It relies on the assumption that the field inside the
``cavity'' is exactly a resonant mode and the field outside the
``cavity'' is exactly the sum of an incoming wave and an outgoing
wave. The actual field satisfying the Maxwell's equations is of course
far more complicated. In this paper, we use a perturbation method to
derive an accurate formula for field enhancement.  
%
%
It reveals that field enhancement depends on the resonant mode,
its coupling with the incident wave, and the frequency of the incident
wave. We pay special attention to the case where the resonant mode has
multiple radiation channels, determine the incident wave that maximize
the field enhancement, and show that field enhancement is
proportional to $\sqrt{Q_{\rm rad}} Q_{\rm dis} /(Q_{\rm rad} + Q_{\rm dis})$ only for special (usually non-optimal) incident waves. 
We also study the  dependence on wavevector for structures supporting
a cBIC, and confirm that the largest field
enhancement is obtained at a wavevector satisfying the 
the critical coupling condition approximately. 

The rest of this paper is organized as follows. In
Sec.~\ref{sec:resonance}, we give some background material on
resonant modes and cBICs in lossy periodic structures, and present some
integral formulas for $Q_{\rm rad}$ and $Q_{\rm dis}$. In
Sec.~\ref{Sec:FieldEnhance}, we drive a rigorous formula for field enhancement by a perturbation
method, find the conditions such that the field enhancement
is proportional to $\sqrt{Q_{\rm rad}} Q_{\rm dis} /(Q_{\rm rad} +
Q_{\rm dis})$, and determine incident waves that maximize the field
enhancement. Numerical examples are presented in
Sec.~\ref{sec:examples} to illustrate how field enhancement depends on
the incident wave and the Bloch wavenumber. 
The paper is concluded with a brief discussion in Sec.~\ref{sec:conclusion}.

\section{Resonant and bound states}
\label{sec:resonance}

We consider a lossy two-dimensional (2D) structure that is invariant
in $z$, symmetric and periodic (with period $L$) in $y$, and
sandwiched between two identical lossless homogeneous 
media given for $x > d$ and $x < -d$, respectively.
The dielectric function satisfies 
\begin{equation}
  \label{refper}
  \varepsilon( {\bf r} ) =\varepsilon(x,y+L) = \varepsilon(x,-y)
\end{equation}
and $\mbox{Im}(\varepsilon) \ge 0$ for all ${\bf r}=(x,y)$,
$\max  [\mbox{Im}(\varepsilon)]  > 0$, and
$\varepsilon({\bf r} ) =\varepsilon_0 \ge 1$ for $ |x| > d$. 
For any $E$-polarized  time-harmonic wave with the time dependence $e^{- i \omega t}$ ($\omega$ is the angular frequency), the $z$-component of the electric field, denoted as $u({\bf r} )$, satisfies the following 2D Helmholtz equation
\begin{equation}
\label{helm} \partial^2_x u + \partial^2_y u  + k^2 \varepsilon({\bf r} ) u = 0,
\end{equation}
where $k = \omega / c$ is the freespace wave number and $c$ is the speed of light in vacuum.


Due to the periodicity, any eigenmode of the structure is a Bloch mode given as
\begin{equation}
  \label{bloch} u({\bf r} ) = \phi({\bf r} ) e^{i \beta y}, 
\end{equation}
where $\phi$ is periodic in $y$ and $\beta$ is the Bloch wavenumber. We
assume $\beta$ is real and $\omega$ (or $k$) is the eigenvalue.
A resonant mode (or resonant state) is an eigenmode satisfying outgoing radiation conditions. 
It radiates out power and diverges as $x\to \pm \infty$.
In contrast, a bound state decays to zero exponentially as $x \to \pm \infty$.
Since the structure is lossy, $\mbox{Im}(\omega)$ is negative for both
resonant and bound states. 
In the surrounding homogeneous media, the Bloch mode can be expanded as
\begin{equation}
\label{fourier} u({\bf r} ) = \sum_{m = -\infty}^{\infty} c^{\pm}_m e^{i [\beta_m y \pm \alpha_m (x \mp d)]}, \quad \pm x > d, 
\end{equation}
where $c_m^{\pm}$, for all integers $m$, are expansion coefficients, $\beta_0=\beta$, and 
\begin{equation}
\label{betaalpha} \beta_m = \beta + 2 m \pi /L, \quad \alpha_m = \sqrt{k^2 \varepsilon_0 - \beta^2_m}. 
\end{equation}
In order for Eq.~(\ref{fourier}) to describe both resonant and bound
states correctly, we need to choose the complex square root in Eq.~(\ref{betaalpha}) to maintain
continuity as $\mbox{Im}(\omega) \to 0$. More precisely, if
$\mbox{Re}(k^2 \varepsilon_0 - \beta_m^2)$ is positive, then
$\alpha_m$ is in the fourth quadrant of the complex plane with
$\mbox{Re}(\alpha_m) > 0$ and $\mbox{Im}(\alpha_m) < 0$, so that $e^{
  i \alpha_m x}$ is an outgoing plane wave (with increasing amplitude)
as $x \to +\infty$; and if $\mbox{Re}(k^2 \varepsilon_0 - \beta_m^2)$
is negative, then $\alpha_m$ is chosen to be in the second quadrant
with $\mbox{Re}(\alpha_m) < 0$ and $\mbox{Im}(\alpha_m) > 0$, so that
$e^{ i \alpha_m x}$ decays exponentially as $x \to +\infty$.  We
denote the set of integers $m$ such that $\mbox{Re}(k^2 \varepsilon_0
- \beta_m^2) > 0$ by $\mathbb{Z}_0$. If $\mathbb{Z}_0$ is not empty
and there is at least one nonzero coefficient $c_m^+$ or $c_m^-$ for an
$m \in \mathbb{Z}_0$, then the eigenmode is a resonant mode that
radiates out power.  If $\mathbb{Z}_0$ is not empty, but $c_m^\pm = 0$
for all $m \in \mathbb{Z}_0$, then the eigenmode is a cBIC. We assume
$\beta$ has been shifted to satisfy $ |\beta| \le
\pi/L$, then a non-empty $\mathbb{Z}_0$ implies that $0 \in
\mathbb{Z}_0$, i.e., $\mbox{Re}(k^2
\varepsilon_0 - \beta^2) > 0$.  If $|\mbox{Im}(\omega) /\omega|$ is
small, this condition can be approximated by $\mbox{Re}(k)
\sqrt{\varepsilon_0} > |\beta|$.

For a resonant mode with a complex frequency $\omega$,
the total $Q$ factor may be defined by  $Q_{\rm tot} = -0.5
\mbox{Re}(\omega) / \mbox{Im}(\omega)$. Meanwhile, according to
Eq.~(\ref{fourier}), we can decompose the field into radiating  and
evanescent components as
\begin{equation}
u({\bf r})  = u_{r} ({\bf r}) + u_{e}({\bf r}), \quad |x| > d, 
\end{equation}
where 
\begin{equation*}
\label{udecay} u_{e}({\bf r}) = \sum\limits_{m \notin
  \mathbb{Z}_0}c^{\pm}_m e^{i [\beta_m y \pm \alpha_m (x \mp d)]},
\quad |x| > d, 
\end{equation*}
and $u_{r} $ is given similarly for $m \in \mathbb{Z}_0$. In 
Appendix A, we prove that 
\begin{equation}
\frac{1}{Q_{\rm tot}} = \frac{1}{Q_{\rm dis}} + \frac{1}{Q_{\rm rad}}, 
\end{equation}
where
\begin{eqnarray}
\label{Qdis}
&& \frac{1}{Q_{\rm dis}} =  \frac{ \mbox{Re}(k^2)  \int_{\Omega_d} \varepsilon_i({\bf
                      r}) |u({\bf r})|^2 d {\bf r}}{ [\mbox{Re}(k)]^2    F}, \\
\label{Qrad}
&& \frac{1}{Q_{\rm rad}} =  \frac{ L \sum_{m \in \mathbb{Z}_0} ( |c_m^-|^2 + |c_m^+|^2) \mbox{Re}(\alpha_m)  }{ [\mbox{Re}(k)]^2 F }, \\
&&  F =  \int_{\Omega_d} \varepsilon_{r}({\bf r}) |u({\bf r})|^2 d {\bf 
  r} + \varepsilon_0 \int_{\Omega_e} |u_{e}({\bf r})|^2 d {\bf r}.
\end{eqnarray}
In the above, $\varepsilon_r({\bf r})$ and $\varepsilon_i({\bf r})$ are the real and imaginary
parts of the dielectric function $\varepsilon({\bf r})$, $\Omega_d$ is
the rectangular domain given by $|x| < d$ and $|y| < L/2$, 
$\Omega_e$ is the union of two semi-infinite strips given by $|x| > d$
and $|y| < L/2$.
The first and second integrals in the expression $F$ are proportional to the energy stored in domain $\Omega_d$ (i.e. the cavity) and the energy of the
evanescent field $u_e$ outside domain $\Omega_d$, respectively. The numerator in Eq.~(\ref{Qdis}) is proportional to the power absorbed by material dissipation. 
The numerator in Eq.~(\ref{Qrad}) is proportional to the power
radiated out as $x \to \pm \infty$. 
The above equations give the well-known
interpretation that $1/Q$ is the ratio of power lost per cycle and
total stored energy, $Q_{\rm dis}$ and
$Q_{\rm rad}$ are quality factors associated with absorption and radiation
losses, respectively~\cite{blio08}. 
The imaginary part of $\omega$ can also be decomposed as
\begin{equation}
\label{ImResFreq} \mbox{Im}(\omega) = \omega''_{\rm dis} + \omega''_{\rm rad}, 
\end{equation}
where $\omega''_{\rm dis}$ and $\omega''_{\rm rad}$ are associated with 
dissipation and radiation losses, respectively, and satisfy 
\begin{equation*}
Q_{\rm dis} = - \frac{\mbox{Re}(\omega)}{2 \omega''_{\rm dis}} \quad
\mbox{and} \quad Q_{\rm rad} = - \frac{\mbox{Re}(\omega)}{2\omega''_{\rm
    rad}}
\end{equation*}

If the resonant mode is scaled such that 
\begin{equation}
\label{normResonant}
\mathop{\mbox{max}}_{ \mathbf{r} \in \Omega_d} |u({\bf r})| = 1,
\end{equation}
then $c_m^{\pm}$ in Eq.~(\ref{fourier}) are dimensionless constants, 
$F$ is proportional to $L^2$, $k$ and $\alpha_m$ are proportional to
$1/L$. Therefore, Eq.~(\ref{Qrad}) leads to 
\begin{equation}
\label{Qc} \frac{1}{Q_{\rm rad}} \sim \frac{1}{\mbox{Re}(k)}
\sum\limits_{m \in \mathbb{Z}_0} \left(  |c_m^-|^2 + |c_m^+|^2 \right)  \mbox{Re}(\alpha_m).
\end{equation}

A cBIC is a special point in a branch of resonant modes. It can be
regarded as a special resonant mode with $Q_{\rm rad} = \infty$ (thus
$Q_{\rm tot} = Q_{\rm dis}$).  Let $\beta_{\diamond}$ be the real
Bloch 
wavenumber of the cBIC, then $Q_{\rm rad}$ of the resonant mode with
Bloch wavenumber $\beta$ tends to infinity as $\beta \to
\beta_{\diamond}$. In the same limit, $Q_{\rm dis}$ of the resonant
mode tends to a finite constant, i.e., the $Q_{\rm dis}$ of the cBIC.

\section{Field enhancement}
\label{Sec:FieldEnhance}

In this section, we analyze field enhancement for incident waves
impinging upon a lossy periodic structure. Using a perturbation 
method, we derive an explicit formula for field enhancement, study
how it depends on the incident waves and the $Q$ factors. In the case of multiple open
diffraction channels, we specify a  
condition on the incident wave for maximum field 
enhancement. It is also shown that when the periodic structure
supports a cBIC, the largest field  
enhancement occurs approximately at the wavenumber satisfying the critical coupling
condition, i.e., $Q_{\rm rad} = Q_{\rm dis}$. 

We consider a 2D lossy symmetric periodic structure as
described in Sec.~\ref{sec:resonance}, assume that there is a resonant
mode $u_*$ with a complex frequency $\omega_*$ depending on the real
Bloch wavenumber $\beta$, and expand $u_*$ as in
Eq.~(\ref{fourier}) with $\alpha_m$ replaced by $\alpha^*_m =
\sqrt{k_*^2 \varepsilon_0   - \beta_m^2}$ for $k_* = \omega_*/c$. 
For the given $\beta$ and a real 
frequency $\omega$ close or equal to $\mbox{Re}(\omega_*)$, we study the
diffraction problem with an incident wave given by 
\begin{equation}
\label{uin} u^{(in)}(\mathbf{r}) = \sum\limits_{m \in \mathbb{Z}_0}
a^{\pm}_m e^{ i [\beta_m y \mp \alpha_m (x \mp d)] }, \quad \pm x > d, 
\end{equation}
where $a_{m}^{\pm}$ (for $m\in \mathbb{Z}_0$)  are the coefficients of
incident plane waves, $\beta_m$ and 
$\alpha_m$ are given  in Eq.~(\ref{betaalpha}),  $\mathbb{Z}_0$ is
the set of all propagating diffraction orders (or open diffraction
channels) and is defined in the previous section. 
 Notice that $u^{(in)}$ consists of incoming plane waves given in the
 left ($x < -d$) and right ($x > d$) sides of the periodic  structure. 
We further assume that the coefficients of the incident
plane waves satisfy
\begin{equation}
\label{amplitude}  \max\left\{ \sum\limits_{m \in \mathbb{Z}_0}  |a_m^-|,  \sum\limits_{m \in \mathbb{Z}_0}  |a_m^+|  \right\} = 1.
\end{equation}
This implies that  the amplitude of $u^{(in)}$ is bounded by  one,
i.e.  $|u^{(in)}(\mathbf{r})| \le  1$ for all $\mathbf{r} \in \Omega_e$. 

Let $u$ be the diffraction solution associated with the incident wave
given above.  If $|\omega-\omega_*|/|\omega_*|$ is small, 
we expand  $u$ in a Laurent series of $\delta =
k-k_*$: 
\begin{equation}
  \label{uexp}
u = \frac{C}{\delta} u_* + u_0 + \delta u_1 + \ldots,
\end{equation}
where $C$ is a constant.  In Appendix B, we show that 
\begin{equation}
\label{C} C = \frac{ i L}{k_* R} \sum_{m \in \mathbb{Z}_0} \left(
  a_m^- c_m^- + a_m^+ c_m^+ \right) \alpha_m^*, 
\end{equation}
where
\[
 R = \int_{\Omega_d} \varepsilon({\bf r})  u_*(x,-y) u_*(\mathbf{r}) d
 {\bf r} + \frac{i L}{2} \sum\limits_{m = -\infty}^{+\infty} \frac{
   (c_m^-)^2   + (c_m^+)^2 }{\alpha^*_m}.
\]
In the above, $c_m^{\pm} $ are the expansion coefficients of $u_*$ as
given in Eq.~(\ref{fourier}). The sum in the right hand side of
Eq.~(\ref{C}) provides the coupling between the resonant mode and the
incident wave. Since the structure has a reflection symmetry in $y$,
the resonant mode reciprocal to $u_*$, with the Bloch factor $e^{- i
  \beta y}$, is simply $u_*(x, -y)$,  and it appears in the expression $R$. 

 The field enhancement factor (or field enhancement), denoted by $\eta$, can be defined as
 the ratio of maximum amplitudes of total field in $\Omega_d$ and
the incident wave in $\Omega_e$, i.e. 
\begin{equation}
  \label{enhancement}
  \eta = \frac{ \mathop{\mbox{max}}\limits_{{\bf r} \in \Omega_d}
    |u({\bf r})| }{ \mathop{\mbox{max}}\limits_{{\bf r} \in \Omega_e}
    |u^{(in)}({\bf r})| }.  
\end{equation}
Since $u_*$ and $u^{(in)}$ are required to satisfy
Eqs.~(\ref{normResonant}) and (\ref{amplitude}), respectively, we have 
\begin{equation}
 \label{etaApprox} 
  \eta \approx { \mathop{\mbox{max}}_{{\bf r} \in
    \Omega_d} |u({\bf r})| } \approx \frac{|C|}{|\delta|}.
\end{equation}
From Eq.~(\ref{C}), it is clear that field enhancement depends
strongly on the incident wave. If the coefficients are chosen such
that $(a^-_m, a^+_m)$ is proportional to $(-c_m^+, c_m^-)$ for each 
$m \in  \mathbb{Z}_0$, then $C = 0 $ and 
there is no enhancement. On the other hand, if
\begin{equation}
  \label{good}
  a_m^\pm  = B \frac{ \overline{\alpha}_m^*  \mbox{Re}( \alpha_m^*) } { |
  \alpha_m^*|^2} \overline{c}_m^\pm,   
\end{equation}
where $B$ is a constant chosen to satisfy Eq.~(\ref{amplitude}), 
$\overline{\alpha}_m^*$ and 
$\overline{c}_m^\pm$ are the complex conjugates of
$\alpha_m^*$ and $c_m^\pm$, respectively, then 
\[
\sum_{m \in \mathbb{Z}_0}
(a_m^- c_m^-   +a_m^+ c_m^+) \alpha^*_m = B \sum_{m \in \mathbb{Z}_0}
( |c_m^-|^2 + |c_m^+|^2) \mbox{Re}(\alpha^*_m).
\]
In that case, using Eq.~(\ref{Qc}) (with $k$ and $\alpha_m$ replaced by $k_*$ and
$\alpha_m^*$, respectively), we obtain 
\begin{equation}
  \label{absC}
  |C| \sim \frac{1}{ L \sqrt{Q_{\rm rad}}}.  
\end{equation}
In addition, if we let $\omega = \mbox{Re}(\omega_*)$, then $\delta =
- \mbox{Im}(k_*) \sim 1/( L Q_{\rm tot})$, and thus
\begin{equation}
\label{etaQ}
\eta  \sim \frac{  Q_{\rm dis} \sqrt{Q_{\rm rad}} }{Q_{\rm rad} + Q_{\rm dis}}.
\end{equation}

Notice that Eq.~(\ref{etaQ}) is useful when 
parameters are introduced so that  $Q_{\rm rad}$ and/or $Q_{\rm
 dis}$ vary rapidly with the parameters, but the resonant mode is nearly invariant. 
For example, Eq.~(\ref{etaQ}) is valid when $\mbox{Im}(\varepsilon)$
is close to zero. As $\mbox{Im}(\varepsilon) \to 0$, $Q_{\rm dis} \to
\infty$ and $\eta \sim \sqrt{Q_{\rm rad}}$. It is
also valid when there is a cBIC and $\beta$ is close to that of the
cBIC. 

If the zeroth diffraction order is the only propagating one, i.e., 
$\mathbb{Z}_0 = \{ 0 \}$, and if $|c_0^-| = 
|c_0^+|$, then formula (\ref{etaQ}) gives the maximum field
enhancement for optimally  chosen incident waves. If 
$\mathbb{Z}_0 \ne  \{ 0 \}$, and if $\mathbb{Z}_0 =  \{ 0 \}$ but 
$|c_0^-| \ne |c_0^+|$, then a larger field enhancement can be obtained
by appropriately choosing the coefficients of the incident wave. 
Let $m_1$ and $m_2$ be integers in $\mathbb{Z}_0$ that attach
the maxima of $ |\alpha_m^* c_m^- |$ and  $|\alpha_m^* c_m^+|$ for all
$m \in \mathbb{Z}_0$, respectively, that is, 
\begin{equation}
  \label{maxc1}
  |\alpha^*_{m_1} c_{m_1}^-|  =
  \max_{ m \in \mathbb{Z}_0} \left|
    \alpha_m^* c_m^-   \right|, \ \
  |\alpha^*_{m_2} c_{m_2}^+|  
  = \max_{ m \in \mathbb{Z}_0} \left| \alpha_m^* c_m^+
 \right|.   
\end{equation}
Under the normalization condition (\ref{amplitude}), the maximum of $|C|$ is
obtained when the 
incident wave in the 
left (for $x < -d$) is a single plane wave of diffraction order $m_1$
with unit magnitude ($|a_{m_1}^-|=1$), 
the incident wave in the right (for $x < d$) is a single plane wave of diffraction order $m_2$
with unit magnitude ($|a_{m_2}^+|=1$), and when 
the phases of $a_{m_1}^-$ and
$a_{m_2}^+$ are chosen such that 
\begin{equation}
  \label{maxc2}
  | a_{m_1}^- c_{m_1}^- \alpha_{m_1}^* +
  a_{m_2}^+ c_{m_2}^+ \alpha_{m_2}^* |
  =
  |  c_{m_1}^- \alpha_{m_1}^* |  +
  | c_{m_2}^+ \alpha_{m_2}^* |. 
\end{equation}
This implies that the maximum of $|C|$ is 
\begin{equation}
\label{cmax}
|C|_{\rm max} = \frac{L}{ |k_* R|}  \left( |  c_{m_1}^- \alpha_{m_1}^* |  +
  | c_{m_2}^+ \alpha_{m_2}^* | \right). 
\end{equation}
The corresponding maximum field enhancement is obtained with the
frequency $\omega = \mbox{Re}( \omega_*)$.

The above theory is derived for the diffraction problem with a fixed
real wavenumber $\beta$.  If there is a cBIC in a band of resonant
modes, it is important to understand how field enhancement 
varies with $\beta$ (or incident angle).
Let $u_\diamond({\bf r})$, $\omega_\diamond$ and $\beta_\diamond$  be
the wave field,  frequency, and wavenumber of the cBIC, respectively.
For the  resonant mode with wavenumber $\beta$, we denote its complex
frequency by $\omega_*(\beta)$, its total $Q$ factor by $Q_{\rm
  tot}(\beta)$, etc. 
As $\beta \to \beta_\diamond$, we have 
$\omega_*(\beta) \to \omega_\diamond$,
$Q_{\rm rad}(\beta) \to \infty$, and $Q_{\rm dis}(\beta) \to Q_{\rm
  dis}(\beta_\diamond) < \infty$. 
If formula~(\ref{etaQ}) [with $\eta$, $Q_{\rm rad}$
and $Q_{\rm dis}$ all depending on $\beta$] is applicable, the maximum
of $\eta(\beta)$ can be found approximately by replacing
$Q_{\rm dis}(\beta)$ by the constant $Q_{\rm dis}(\beta_\diamond)$. This gives rise to a  wavenumber
  $\beta_c$, such that
  $Q_{\rm rad}(\beta_c) =
Q_{\rm dis}(\beta_c) \approx Q_{\rm dis}(\beta_\diamond)$.
This is the so-called critical coupling condition.
The field enhancement is maximized
approximately at this critical wavenumber $\beta_c$, and 
  \begin{equation}
  \label{etaQ2}
  \eta(\beta_c) \sim [ Q_{\rm dis}(\beta_\diamond) ]^{1/2}
    \sim  \frac{L}{ \left[  \int_{\Omega_d} \varepsilon_i ({\bf r}) 
        |u_\diamond({\bf r})|^2 d {\bf r} \right]^{1/2} }
  \end{equation}
It should be emphasized that, strictly speaking,  the above is only valid for 
  incident waves with frequency
  $\omega = \mbox{Re}[ \omega_*(\beta_c)] $
  and with coefficients satisfying Eq.~(\ref{good}).

 \section{Numerical examples}
 \label{sec:examples}

\begin{figure}[htp]
\includegraphics[scale=0.8]{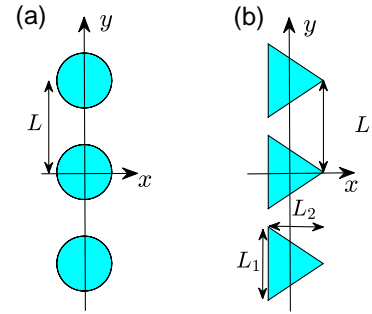}
\caption{(a): A periodic array of circular cylinders with radius $a$. (b): A periodic array of triangular cylinders with
  base $L_1$ and height $L_2$.}
\label{figArrayCylinders}
\end{figure}

\begin{figure}[htp]
\includegraphics[scale=0.66]{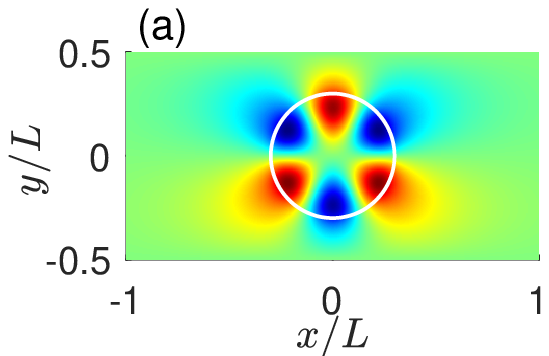}
\includegraphics[scale=0.66]{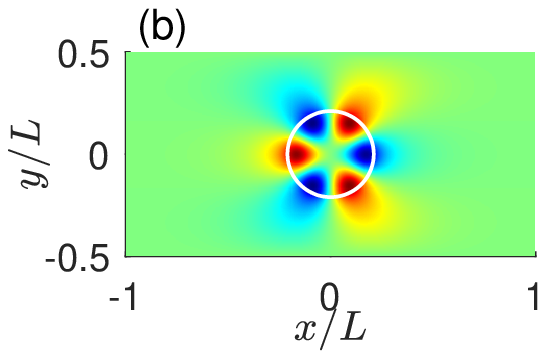}
\includegraphics[scale=0.66]{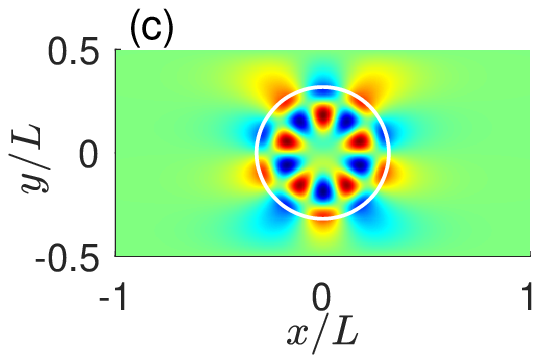}
\includegraphics[scale=0.66]{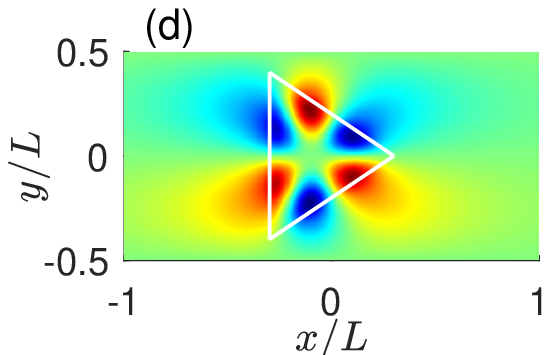}
\caption{Field patterns ($\mbox{Re}(u)$) of the four cBICs listed in Table~\ref{CSPBIC}. (a)-(d): cBICs 1 to 4, respectively.}
\label{figCSPBIC}
\end{figure}

 To validate the analytic results derived in the previous section, we
 present a few numerical examples for 
 diffraction problems of periodic arrays of cylinders with single and
 multiple propagating  diffraction orders.  As shown in
 Fig.~\ref{figArrayCylinders}, 
 the periodic arrays consist
 of circular or triangular cylinders (with dielectric 
 constant $\varepsilon_1$) surrounded by air (with dielectric constant
 $\varepsilon_0 = 1$). The radius of the circular cylinders is $a$. The cross section
 of a triangular cylinder is an isosceles triangle with base $L_1 =
 0.8L$ and height $L_2 = 0.6L$, where $L$ is the period in $y$.
 In Table~\ref{CSPBIC},
\begin{table}[htp]
\centering 
\begin{tabular}{c|c|c|c|c} \hline 
cBIC & $a/L$  & $\beta_{\diamond}$  & ${\omega_{\diamond} L}/(2\pi c)$  & $\mathbb{Z}_0$
  \\ \hline  
 1 & $0.3 $  & $0$ & $ 0.7718567- 3.03 \times 10^{-6} i$ & $\{ 0 \} $   \\ \hline 
 2 & $0.2017 $  & $\pi/L$ & $ 1.0935834-4.25 \times 10^{-6} i$ & $\{ -1, 0 \}$   \\ \hline 
 3 & $0.3194$ &  $0$ & $ 1.5717916 - 6.29 \times 10^{-6} i$ & $\{ -1,0,1\}$  \\ \hline 
4 & {\rm N.A.} & $0$ & $0.8971846 - 3.40 \times 10^{-6} i $ & $\{0\}$ \\ \hline 
\end{tabular}
\caption{Complex BICs in lossy periodic arrays of cylinders with $\varepsilon_1 = 12 + 0.0001 i$.
  The first three cBICs exist in the array of circular cylinders with
  radius $a$. The fourth cBIC exists in the the array of triangular
  cylinders. $\beta_{\diamond}$ and $\omega_{\diamond}$ are the Bloch
  wavenumber and complex frequency of the cBICs. $\mathbb{Z}_0$ is the
  integer set for open diffraction channels.} 
  \label{CSPBIC}
\end{table}
we list four cBICs for the periodic arrays with $\varepsilon_1  = 12 + 0.0001 i$. 
The first three and the fourth cBICs exist in the periodic arrays of
circular and triangular cylinders, respectively. The second and third
cBICs have multiple open radiation channels ($\mathbb{Z}_0 \ne \{
0 \}$),  and are found by tuning the radius of 
the cylinders \cite{bulg14}.  The first and fourth cBIC are fully
symmetry-protected. 
The field patterns of these four cBICs are shown in Fig.~\ref{figCSPBIC}. 

\begin{figure}[htp]
\centering 
\includegraphics[scale=0.7]{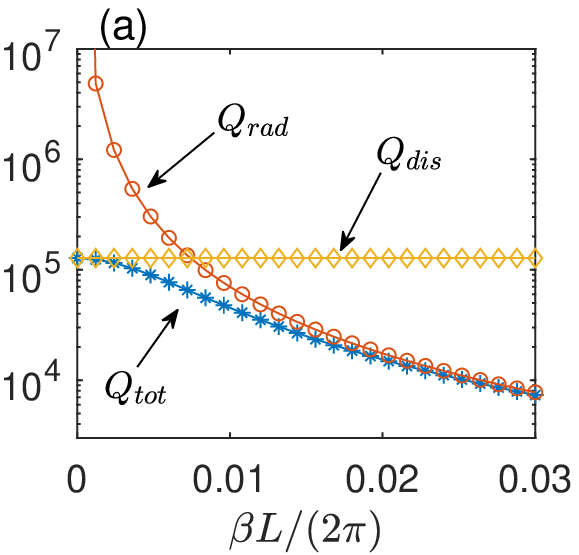}
\includegraphics[scale=0.7]{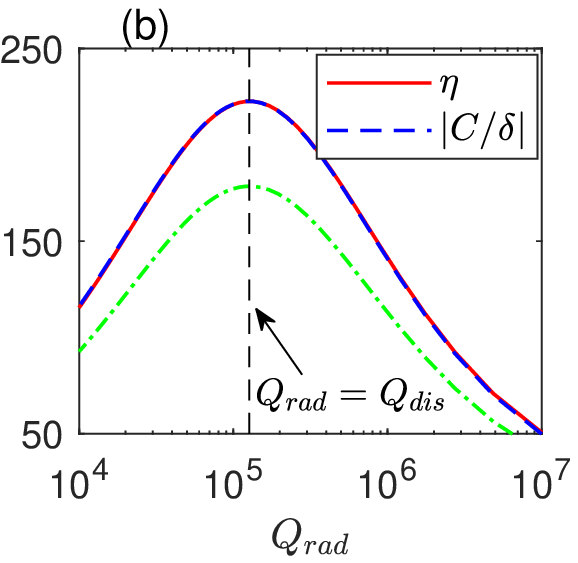}
\includegraphics[scale=0.69]{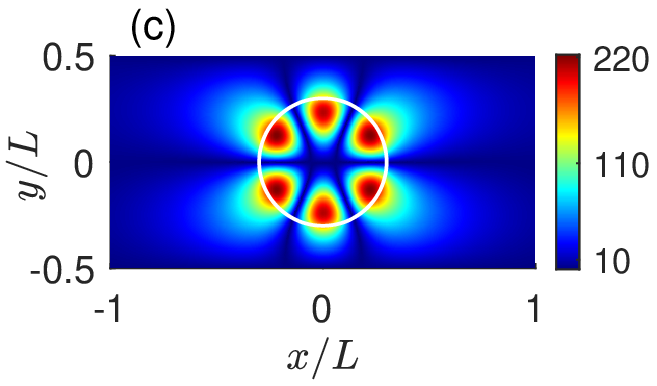}
\includegraphics[scale=0.69]{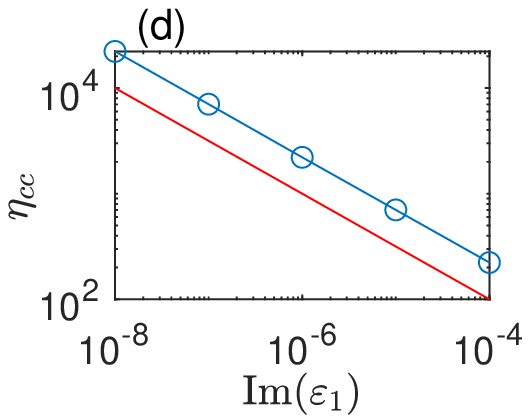}
\caption{Field enhancement near cBIC 1. (a): $Q_{\rm
    tot}$, $Q_{\rm rad}$ and $ Q_{\rm dis}$ of the resonant mode as functions of $\beta$. (b): Field enhancement $\eta$
  (red solid curve), $|C/\delta|$ (blue dashed curve), and 
  $\sqrt{Q_{\rm rad}} Q_{\rm dis} /( Q_{\rm rad} +  Q_{\rm dis})$
(green dash-dot  curve), as functions of $Q_{\rm rad}$. 
(c): Magnitude of the diffraction solution at critical coupling, i.e.,
$|u|$ for $\beta = \beta_c$.
(d) Field enhancement at critical coupling $\eta(\beta_c)$  (``$\circ$'')  for
different values of $\mbox{Im}(\varepsilon_1)$. The solid red line is
the reference line for $ 1/\sqrt{\mbox{Im}(\varepsilon_1)}$.}  
\label{fig:Example1}
\end{figure}

For cBIC 1, the zeroth diffraction order is the only open
radiation channel. Since the structure is symmetric in
$x$, the radiation channels in the left and right sides can be
considered as the same. In Fig.~\ref{fig:Example1}(a),
we show
$Q_{\rm  tot}$, $Q_{\rm dis}$ and $Q_{\rm rad}$ for the resonant
mode  near the cBIC as  functions of $\beta$. 
As $\beta \to \beta_\diamond=0$, $Q_{\rm rad} \to \infty$,  and
$Q_{\rm dis}$ converges to a constant ($\approx 1.27 \times 10^5$).
The critical coupling condition is satisfied at Bloch wavenumber $\beta_c
\approx 0.0074 
(2\pi/L)$. The corresponding resonant  
frequency is $\omega_* \approx 0.7718519 - 6.05 \times 10^{-6} i (2\pi
c/L)$.  Since cBIC 1 is even in $x$, the nearby resonant mode is also
even in $x$, and thus $c_0^+ = c_0^-$. To maximize field enhancement,
we choose an incident wave with frequency $\omega =
\mbox{Re}(\omega_*)$ and coefficients $a_{0}^+ = a_0^- = 1$. 
Using the numerical solutions for different $\beta$, we evaluate the 
field enhancement $\eta$ and its approximation $|C/\delta|$ following 
Eqs.~(\ref{enhancement}) and (\ref{C}), respectively. 
The computed $\eta$,  $|C/\delta|$ and  
$ \sqrt{Q_{\rm rad}} Q_{\rm dis} /( Q_{\rm rad} +  Q_{\rm dis})$ 
 are shown  in Fig.~\ref{fig:Example1}(b) as
functions of $Q_{\rm rad}$ (which is related to $\beta$).  It can be seen
that $\eta$ is well approximated by  $|C/\delta|$, and is indeed
proportional to  $ \sqrt{Q_{\rm rad}} Q_{\rm dis} /( Q_{\rm rad} +  Q_{\rm dis})$. 
As indicated by the vertical line in Fig.~\ref{fig:Example1}(b), 
the field enhancement reaches the maximum $\eta(\beta_c)  \approx 223$ when
the critical coupling condition is satisfied. 
The corresponding diffraction solution is shown in
Fig.~\ref{fig:Example1}(c). The above numerical results are obtained for the
fixed $\varepsilon_1 = 12 + 0.0001i$. Equation~(\ref{etaQ2}) indicates
that field enhancement  is proportional to
$1/\sqrt{\mbox{Im}(\varepsilon_1)}$,  if $\mbox{Im}(\varepsilon_1)$ is
a constant. To validate this result, we keep
$\mbox{Re}(\varepsilon_1)$ fixed and calculate $\eta(\beta_c)$ for
different values of $\mbox{Im}(\varepsilon_1)$. The results are shown
in Fig.~\ref{fig:Example1}(d).

\begin{figure}[htp]
\centering 
\includegraphics[scale=0.7]{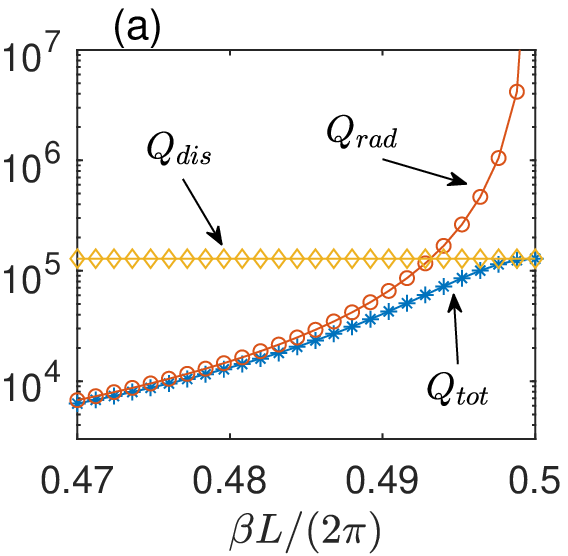}
\includegraphics[scale=0.7]{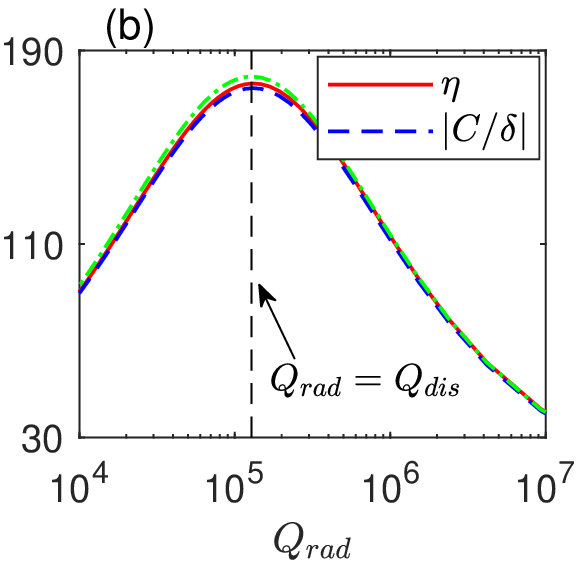}
\includegraphics[scale=0.69]{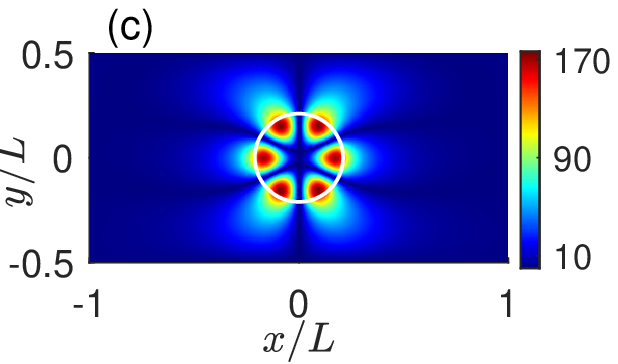}
\includegraphics[scale=0.7]{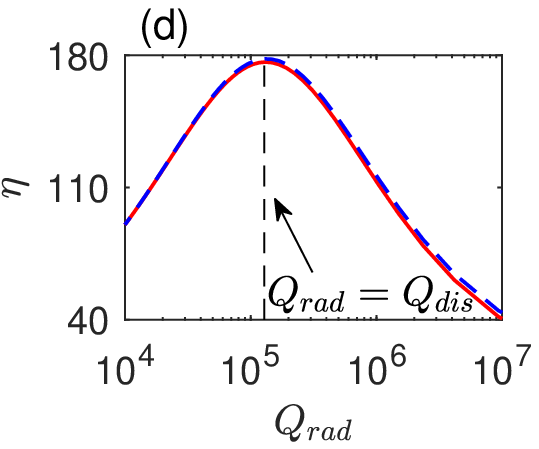}
\caption{Field enhancement near cBIC 2. (a): $Q_{\rm tot}$,
  $Q_{\rm rad}$ and $ Q_{\rm dis}$ of the resonant mode as functions
  of $\beta$. (b):  Field 
  enhancement $\eta$ (red solid curve), $|C/\delta|$ (blue dashed
  curve),   and $\sqrt{Q_{\rm rad}}  Q_{\rm dis}/(Q_{\rm rad} + Q_{\rm dis})$
  (green dash-dot curve),  as functions of $Q_{\rm rad}$.
  (c): Magnitude of the diffraction
  solution at critical coupling, i.e., 
  $|u|$ for $\beta = \beta_c$. 
(d): Field enhancement $\eta(\beta)$ as functions of $Q_{\rm rad}$,  for
incident waves with 
coefficients given by Eq.~(\ref{good}) (red solid curve) and
Eq.~(\ref{bic2coef1}) (blue dashed curve), respectively.}
\label{fig:Example2}
\end{figure}

For cBIC 2, there are two radiation channels (diffraction orders $m=0$ and $-1$) in each side of the
periodic array. Due to the reflection symmetry in $x$, the radiation channels in the
left and right sides are  identical. The symmetry in $y$ is also
important to this cBIC. If a resonant mode  is even
in $y$ and has a Bloch wavenumber $\beta = \pi/L$, then $c_0^\pm =
c_{-1}^\pm$. For the array of circular cylinders, such a
resonant mode with $\beta = \pi/L$ becomes a cBIC (with $c_0^\pm =
c_{-1}^\pm=0$), if the radius is tuned to $a = 0.2017L$. 
For this $a$, we consider resonant modes with $\beta$ near
$\beta_\diamond=\pi/L$, 
and show $Q_{\rm tot}$, $Q_{\rm dis}$ and $Q_{\rm rad}$ as  functions
of $\beta$ in Fig.~\ref{fig:Example2}(a).
As $\beta \to \pi/L$, $Q_{\rm dis}$ converges to a constant ($\approx
1.28 \times 10^5$). The critical coupling condition is satisfied at the
Bloch wavenumber $\beta_c \approx 0.4932 (2\pi/L)$. The corresponding resonant  
frequency is $\omega_* \approx 1.0935827 - 8.53 \times 10^{-6} i (2\pi
c/L)$.
For an incident wave with frequency $\omega = \mbox{Re}(\omega_*)$ and
coefficients given by Eq.~(\ref{good}), we 
solve the diffraction problem, and show field enhancement
$\eta$, its approximation $|C/\delta|$ and $ \sqrt{Q_{\rm rad}}
Q_{\rm dis} /( Q_{\rm rad} +  Q_{\rm dis})$, as functions of $Q_{rad}$,  in
Fig.~\ref{fig:Example2}(b). 
For $\beta = \beta_c$, field enhancement
reaches its maximum $\eta(\beta_c) \approx 176$.
The corresponding diffraction solution (i.e. $|u|$) is shown in
Fig.~\ref{fig:Example2}(c).
For $\beta=\beta_c$, the radiation coefficients $c^+_0$ and $c^+_{-1}$
satisfy $|c^+_{-1} \alpha^*_{-1}| / | c^+_{0} \alpha^*_0 | \approx
0.97$. Therefore, according to the theory developed in Sec.~III, a
larger field enhancement can be obtained if the incident wave is  
concentrated in the zeroth diffraction order. For example, if the
coefficients of the incident wave are given by 
\begin{equation}
  \label{bic2coef1}
(a^+_0, a^+_{-1}) = - (a^-_0, a^-_{-1})    = (1, 0),  
\end{equation}
then $\eta(\beta_c) \approx 179 > 176$.
In Fig.~\ref{fig:Example2}(d), we compare the field enhancement $\eta$
as a function of $Q_{\rm rad}$ (related to $\beta$), for incident
waves with coefficients given by Eqs.~(\ref{good}) and
(\ref{bic2coef1}), respectively.

The third cBIC exist in the frequency range with three open radiation
channels (diffraction orders $m=0$ and $\pm 1$) in each side of the
structure.  The reflection symmetry in $x$ ensures that the 
radiation channels in the left and right are identical.
The reflection symmetry in $y$ is again important to the existence of 
cBIC 3. Notice that the electric field of cBIC 3 is odd in $y$. In any periodic array of circular
cylinders, a resonant mode
with $\beta = 0$ and an odd-in-$y$ field must have $c_0^\pm = 0$ and
$c_1^\pm = - c_{-1}^\pm$.
For the given $\varepsilon_1$, as $a$ is tuned to $0.3194L$, the
resonant mode becomes cBIC 3 with $c_1^\pm = - c_{-1}^\pm = 0$. 

\begin{figure}[htp]
\centering 
\includegraphics[scale=0.7]{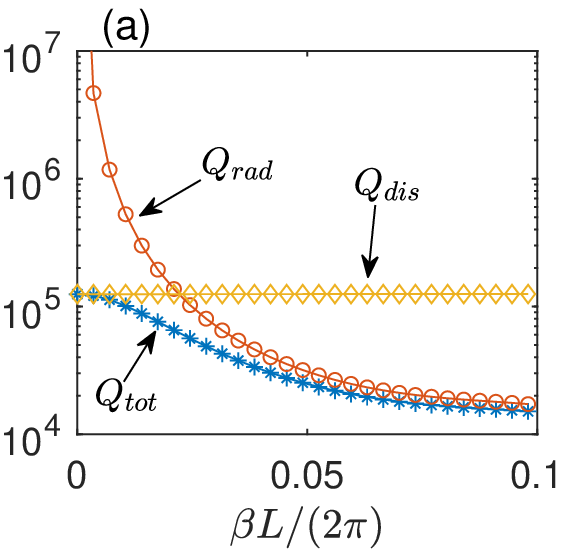}
\includegraphics[scale=0.7]{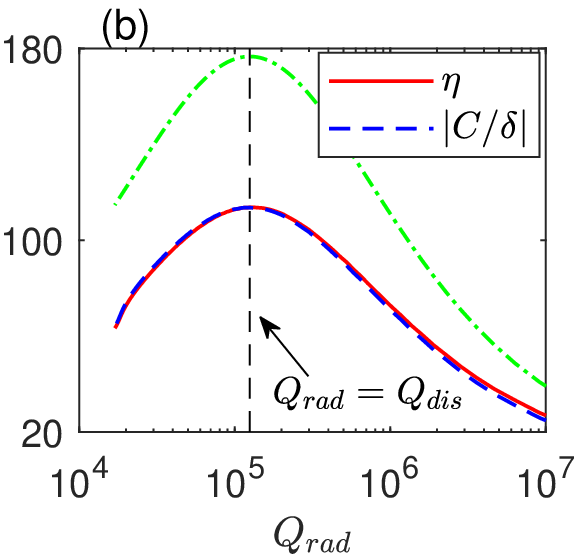}
\includegraphics[scale=0.65]{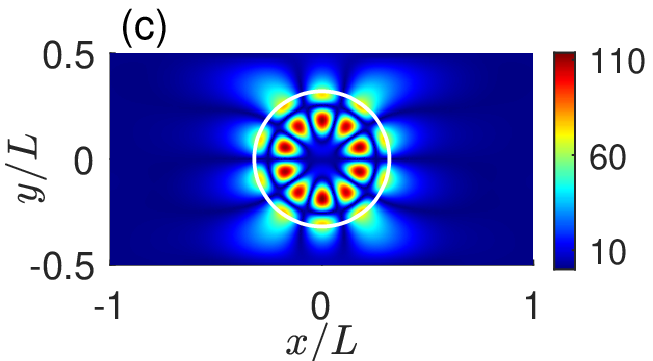}
\includegraphics[scale=0.7]{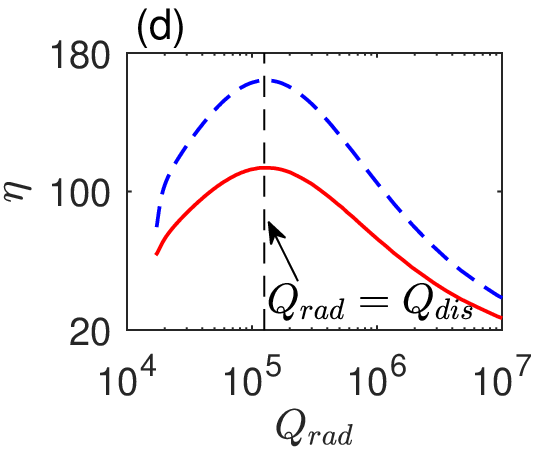}
\caption{Field enhancement near cBIC 3. (a): $Q_{\rm tot}$, $Q_{\rm   rad}$ and $ Q_{\rm dis}$ of the resonant mode as functions of
  $\beta$. (b):   Field enhancement $\eta$ (red solid curve), $|C/\delta|$ (blue dashed 
  curve), and $\sqrt{Q_{\rm rad}}  Q_{\rm dis}/(Q_{\rm rad} + Q_{\rm dis})$
  (green dash-dot curve),  as functions of $Q_{\rm rad}$.
  (c): Magnitude of the diffraction solution at  critical
  coupling, i.e. $|u|$ for $\beta = \beta_c$. 
  (d): Field enhancement $\eta(\beta)$ as functions of
  $Q_{\rm rad}$, for incident waves with coefficients given by
  Eq.~(\ref{good})  (red solid curve) and Eq.~(\ref{bic3coef1}) (blue
  dashed curve), respectively.}
\label{fig:Example3}
\end{figure}

Like the previous two examples, we calculate the resonant modes near
cBIC 3 for $\beta$ near $\beta_\diamond  =0$, and show $Q_{\rm tot}$,
$Q_{\rm rad}$ and $Q_{\rm dis}$ in Fig.~\ref{fig:Example3}(a). 
The limit of $Q_{\rm dis}$ (as $\beta \to 0$) is approximately $1.25
\times 10^5$. The critical coupling condition is satisfied at Bloch
wavenumber $\beta_c \approx 0.0221 (2\pi/L)$, and the corresponding resonant  
frequency is $\omega_*(\beta_c) \approx 1.5718020 - 1.26 \times 10^{-5} i (2\pi
c/L)$.  For each $\beta$, we solve the diffraction problem for an incident wave with
frequency $\omega = \mbox{Re}[ \omega_* (\beta) ]$ and coefficients
given by Eq.~(\ref{good}), and show field enhancement $\eta$, its
approximation $|C/\delta|$, and $\sqrt{Q_{\rm rad}} Q_{\rm dis} /( Q_{\rm rad} +  Q_{\rm dis})$, in
Fig.~\ref{fig:Example3}(b). 
The maximum field enhancement is $\eta(\beta_c) \approx 114$.
The corresponding diffraction solution is shown in
Fig.~\ref{fig:Example3}(c).
%
%
For $\beta = \beta_c$, the radiation coefficients of the resonant mode 
satisfy $ |\alpha_0^* c_0^+ | : |\alpha_{-1}^*
c_{-1}^-| :  |\alpha_{1}^* c_{1}^-| \approx 1 : 0.35 :
0.34$.
Therefore, a larger field enhancement can be obtained, if incident wave 
consists of a single plane wave in the zeroth diffraction order in
each side of the periodic array.  
For the incident wave with coefficients
\begin{equation}
  \label{bic3coef1}
 (a^+_0, a^+_{-1}, a^+_1) = (a^-_0, a^-_{-1}, a^{-}_1)  = (1, 0, 0), 
\end{equation}
we obtain the  field enhancement $\eta(\beta_c) \approx 164 >
114$.
%
To see the difference for $\beta \ne \beta_c$, we compare the
field enhancement $\eta(\beta)$ for incident waves given by Eqs.~(\ref{good}) and
(\ref{bic3coef1}), respectively,  in Fig.~\ref{fig:Example3}(d).

\begin{figure}[htp]
\centering 
\includegraphics[scale=0.7]{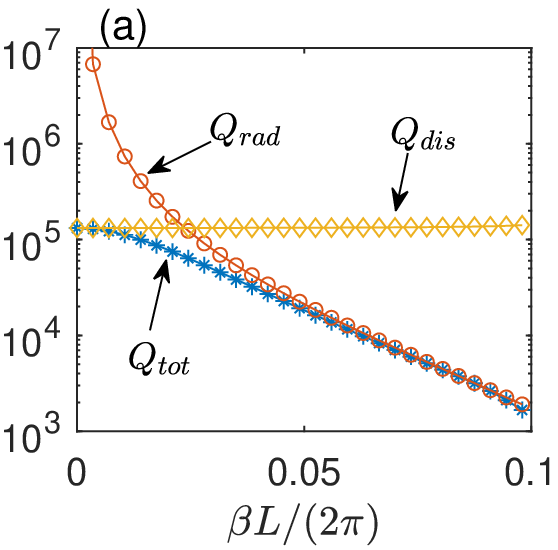}
\includegraphics[scale=0.7]{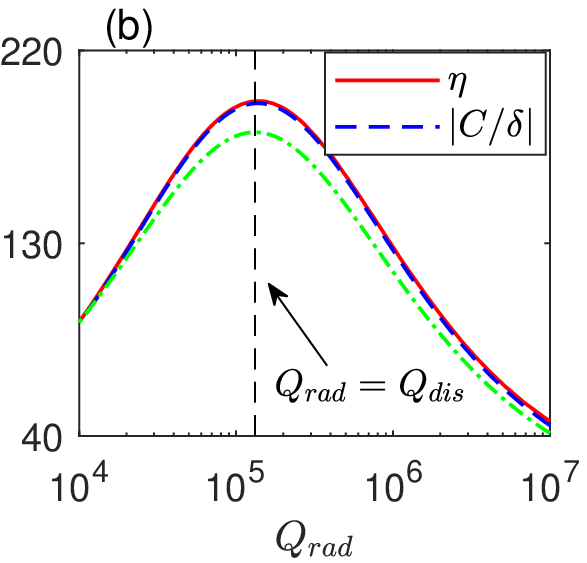}
\includegraphics[scale=0.65]{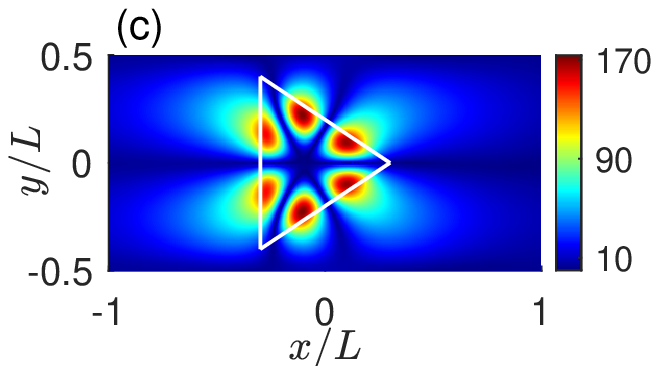}
\includegraphics[scale=0.69]{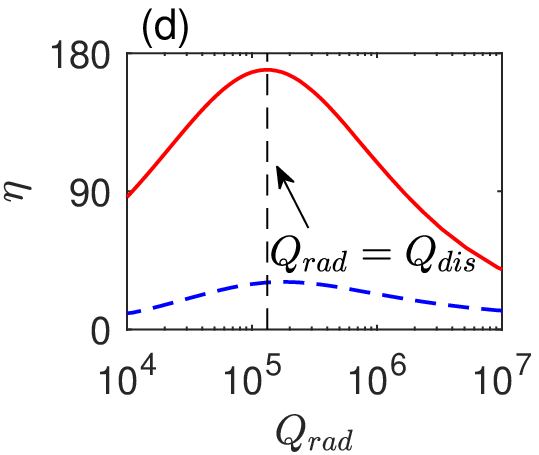}
\caption{Field enhancement near cBIC 4. (a) $Q_{\rm tot}$, $Q_{\rm
    rad}$ and $ Q_{\rm dis}$ of the resonant mode as functions of
  $\beta$.  (b):  Field enhancement $\eta$ (red solid curve) and 
  $|C/\delta|$ (blue dashed curve) for the incident wave with coefficients
  satisfying Eq.~(\ref{bic4coef1}),  and $\sqrt{Q_{\rm rad}}
  Q_{\rm dis}/(Q_{\rm rad} + Q_{\rm dis})$  (green dash-dot curve),
  as functions of $Q_{\rm rad}$.   
  (c): Magnitude of the diffraction solution at critical coupling,
  i.e., $|u|$ for $\beta = \beta_c$. 
(d): Field enhancement $\eta$   as functions of   $Q_{\rm rad}$, for
incident waves with coefficients $(a_0^+, a_0^-) = 
  (1, 0)$ (red solid curve) and $(0, 1)$ (blue dashed curve), respectively.}
\label{fig:Example4}
\end{figure}

The fourth cBIC has only a single radiation channel (diffraction order
$m=0$) in each side of the structure, but the left and right channels
are different due to the lack of symmetry in $x$. 
In Fig.~\ref{fig:Example4}(a),
we show $Q_{\rm tot}$, $Q_{\rm dis}$ and $Q_{\rm rad}$ as functions of
$\beta$ for resonant modes near cBIC 4.
The limit of $Q_{\rm dis}$,
as $\beta \to \beta_\diamond=0$, is approximately $1.32 \times 10^5$.
The critical coupling condition is satisfied as $\beta_c \approx 0.0237
(2\pi/L)$. The corresponding resonant frequency is $\omega_* \approx
0.8970943 - 6.77 \times 10^{-6} i (2\pi c/L)$, and the left and right
radiation coefficients satisfy $|c_0^-/c_0^+| \approx 0.15$.
Since there is only one propagating diffraction order,
Eq.~(\ref{good}) implies that $(a_0^+, a_0^-)$ is proportional to $(\overline{c}_0^+, 
\overline{c}_0^-)$, but a larger field enhancement can be achieved if
$a_0^\pm$ satisfy 
\begin{equation}
  \label{bic4coef1}
  |a_0^\pm | = 1,  \quad |a_0^- c_0^- + a_0^+ c_0^+ |
  = |c_0^-|+|c_0^+|.  
\end{equation}
%
In Fig.~\ref{fig:Example4}(b), we show field enhancement
$\eta(\beta)$, its approximation $|C/\delta|$, and 
$ \sqrt{Q_{\rm rad}} Q_{\rm dis}/(Q_{\rm rad} + Q_{\rm dis})$, 
as functions of $Q_{\rm rad}$, for an incident wave with frequency 
$\omega = \mbox{Re} [ \omega_*(\beta) ] $ and coefficients
satisfying Eq.~(\ref{bic4coef1}). It can be seen that $\eta$ is well
approximated by $|C/\delta|$, but is not proportional to 
$ \sqrt{Q_{\rm rad}} Q_{\rm dis}/(Q_{\rm rad} + Q_{\rm dis})$. 
 The maximum field enhancement is
approximately $\eta(\beta_c)  = 195$.  The corresponding 
diffraction solution is shown in Fig.~\ref{fig:Example4}(c).
%
%
In Fig.~\ref{fig:Example4}(d), we show the field enhancement
$\eta(\beta)$ one-sided incident waves with coefficients $(a_0^+,
a_0^-) = (1, 0)$
(red solid curve) 
and $(0, 1)$ (blue dashed curve), respectively. 
The field enhancement is clearly dominated by the right incident wave. 

\section{Conclusion}
\label{sec:conclusion}

In this paper, we studied resonant field enhancement for diffraction
problems of 2D lossy
periodic structures based on a rigorous perturbation method. The
approximation $\eta \approx |C/\delta|$
is valid for any incident wave with any frequency $\omega$ near the complex
resonant frequency $\omega_*$, and is highly accurate as shown in the numerical
examples. Equation~(\ref{etaQ}) is a simple scaling law between the field
enhancement $\eta$ and the radiation and dissipation $Q$ factors, 
but if the resonant mode has multiple radiation channels, 
Eq.~(\ref{etaQ}) is only valid for specific incident waves with coefficients given by
Eq.~(\ref{good}) and a frequency $\omega$ exactly equals to  the real
part of $\omega_*$, and it does not give the largest field
enhancement. In fact,  we showed that the largest field
enhancement is obtained when the incident wave
consists of a single plane wave in each side of the periodic structure
with coefficients satisfying Eqs.~(\ref{maxc1}) and (\ref{maxc2}). We
also analyzed the dependence of field enhancement $\eta$ on Bloch
wavenumber $\beta$, assuming the periodic structure supports a
cBIC. Consistent with existing studies, the field enhancement is 
approximately maximized at the wavenumber $\beta_c$ such that radiation and
dissipation $Q$ factors are equal (the critical coupling
condition).

Our theoretical results  should be useful in applications where a large field
enhancement is important, especially when the related resonant mode
has multiple radiation channels. Although our theory is developed
only for structures with a one-dimensional periodicity, it can be extended to structures
with two-dimensional periodicity.
%

\section*{Acknowledgement}
The authors acknowledge support from the Natural Science Foundation of Chongqing, 
China (Grant No. cstc2019jcyj-msxmX0717),  
and the Research Grants Council of Hong Kong Special Administrative Region, China (Grant No. CityU 11305518). 

\section*{Appendix A}
\renewcommand{\theequation}{A\arabic{equation}}
\setcounter{equation}{0}

Let $u$ be a resonant mode with a real Bloch wavenumber $\beta$ and
a complex resonant frequency $\omega$.  
Multiplying  both sides of Eq.~(\ref{helm}) by $\overline{u}$ and
integrating the result on domain $\Omega_{d}$, we have 
\begin{equation}
\label{A1}   \int_{\partial \Omega_d} \overline{u} \frac{\partial u}{\partial \nu} ds - \int_{\Omega_d} \nabla \overline{u} \cdot \nabla u d {\bf r} + k^2 \int_{\Omega_d} \varepsilon({\bf r}) |u|^2 d {\bf r} = 0, 
\end{equation}
where $\partial \Omega_d$ is the boundary of $\Omega_d$ and $\nu$ is
the outward unit normal vector of $\Omega_d$. In Appendix A of
Ref.~\cite{hu20_1}, it is proved that 
\begin{equation}
\label{A2}   \int_{\partial \Omega_d} \overline{u} \frac{\partial u}{\partial \nu} ds = i L \sum_{m=-\infty}^{\infty} \alpha_m ( |c_m^+|^2 + |c_m^-|^2 ).  
\end{equation}
Taking the imaginary parts of Eqs.~(\ref{A1}) and (\ref{A2}), we
obtain 
\begin{eqnarray}
\label{A3} & &  L \sum_{m=-\infty}^{\infty} \mbox{Re}(\alpha_m) ( |c_m^+|^2 + |c_m^-|^2 ) + \mbox{Re}(k^2) \int_{\Omega_d} \varepsilon_i({\bf r}) |u|^2 d {\bf r} \nonumber \\
 & & + \mbox{Im}(k^2) \int_{\Omega_d} \varepsilon_r({\bf r}) |u|^2 d {\bf r} =0.
\end{eqnarray}
As shown in Appendix A of Ref.~\cite{hu20_1}, the above leads to 
$$  L \sum_{m \notin \mathbb{Z}_0} \mbox{Re}(\alpha_m) ( |c_m^+|^2 +
|c_m^-|^2 ) = \mbox{Im}(k^2) \varepsilon_0 \int_{\Omega_e} |u_e|^2 d
{\bf r}. $$ Therefore, Eq.~(\ref{A3}) gives rise to 
\begin{eqnarray}
\label{A4}   \mbox{Im}(k^2) &=& - \frac{ \mbox{Re}(k^2) \int_{\Omega_d} \varepsilon_i({\bf r}) |u|^2 d {\bf r} }{F}  \nonumber \\ 
   &-&   \frac{L \sum_{m \in \mathbb{Z}_0} \mbox{Re}(\alpha_m) ( |c_m^+|^2 + |c_m^-|^2 ) }{F},
\end{eqnarray}
where $F$ is defined in Sec.~\ref{sec:resonance}. Notice that
$$\mbox{Im}(k^2) = 2 \mbox{Re}(k) \mbox{Im}(k),$$
we have
\begin{eqnarray}
  \label{A5} \frac{1}{Q_{tot}} &=& - \frac{\mbox{Im}(k^2)}{ \left[ \mbox{Re}(k) \right]^2} =  \frac{ \mbox{Re}(k^2) \int_{\Omega_d} \varepsilon_i({\bf r}) |u|^2 d {\bf r} }{\left[ \mbox{Re}(k) \right]^2 F}  \nonumber \\ 
     &-&   \frac{L \sum_{m \in \mathbb{Z}_0} \mbox{Re}(\alpha_m) ( |c_m^+|^2 + |c_m^-|^2 ) }{\left[ \mbox{Re}(k) \right]^2 F}.
  \end{eqnarray}
The first and second terms on the right hand side of Eq.~(\ref{A5}) gives to Eqs.~(\ref{Qdis}) and (\ref{Qrad}), respectively.

\section*{Appendix B}
\renewcommand{\theequation}{B\arabic{equation}}
\setcounter{equation}{0}

Let $u_*$ be a resonant mode with a real Bloch wavenumber $\beta$ and
a complex frequency $\omega_*$, and be expanded as in
Eq.~(\ref{fourier}) with $\alpha_m$ replaced by $\alpha^*_m =
\sqrt{k_*^2 \varepsilon_0 - \beta_m^2}$  for $k_* = \omega_*/c$.
Since the structure has a reflection symmetry in $y$, $v_* = u_*(x,-y)$ is a resonant 
mode for  Bloch wavenumber $-\beta$ and the same complex frequency
$\omega_*$, and 
$$ v_* = u_*(x,-y) =  \sum_{m = -\infty}^{\infty} c^{\pm}_m e^{i [-\beta_m y \pm \alpha^*_m (x \mp d)]}.$$
For the incident wave $u^{(in)}$ given in Eq.~(\ref{uin}), the
diffraction solution $u$ satisfies Eq.~(\ref{helm}) and
boundary condition
\begin{equation}
  \label{BC}
\pm \frac{\partial u}{\partial x} = \mathcal{B} u - 2 i \sum_{m \in
  \mathbb{Z}_0} \alpha_m a_{m}^{\pm} e^{i \beta_m y}, \quad x = \pm d,
\end{equation}
where $\mathcal{B}$ is the operator defined in Eq.~(24) of
Ref.~\cite{hu20_1}. Inserting expansion (\ref{uexp}) into
Eq.~(\ref{helm}) and boundary condition (\ref{BC}), and comparing the
coefficient of $\delta^j$ for $j \geq -1$, we obtain the equation 
\begin{equation}
  \label{Equ0}
\partial_x^2 u_0 + \partial_y^2 u_0 + k_*^2 \varepsilon u_0 = -2 C k_* \varepsilon u_*
\end{equation}
and boundary condition
\begin{equation*}
  \label{BCu0}
\pm \frac{\partial u_0}{\partial x} - \mathcal{B}_* u_0 = C \mathcal{B}_1 u_* - 2 i \sum_{m \in \mathbb{Z}_0} \alpha^*_m a_{m}^{\pm} e^{i \beta_m y}, \quad x = \pm d
\end{equation*}
for $u_0$, where $\mathcal{B}_*$ and $\mathcal{B}_1$ are operators
defined in Ref.~\cite{hu20_1}.  

Multiplying Eq.~(\ref{Equ0}) by $v_*$ and integrating the result on
domain $\Omega_d$, we have 
\begin{equation}
  \label{B3}
   \int_{\partial \Omega_d} \left( v_* \frac{\partial u_0}{\partial \nu} - u_0 \frac{\partial v_*}{\partial \nu} \right) ds = -2 C k_* \int_{\Omega_d} \varepsilon v_*  u_* d {\bf r}. 
\end{equation}
Following the procedure given in Appendix B of Ref.~\cite{hu20_1}, we
can show that 
\begin{eqnarray}
  \label{B4}
   && \int_{\partial \Omega_d} \left( v_* \frac{\partial u_0}{\partial \nu} - u_0 \frac{\partial v_*}{\partial \nu} \right) ds + 2 i L \sum_{m \in \mathbb{Z}_0} (a_m^+ c_m^+ + a_m^- c_m^-) \alpha_m^* \nonumber \\
    &&= i L k_* C \sum_{m = -\infty}^{\infty} \frac{(c_m^+)^2 + (c_m^-)^2}{\alpha_m^*}.
\end{eqnarray}
Inserting Eq.~(\ref{B3}) into Eq.~(\ref{B4}) and noticing  that $v_* =
u_*(x,-y)$, we obtain Eq.~(\ref{C}).

\end{document}